# Room-temperature Highly-Tunable Coercivity and Highly-Efficient Nonvolatile Multi-States Magnetization Switching by Small Current in Single 2D Ferromagnet $Fe_3GaTe_2$

*Gaojie Zhang*[1,2], *Hao Wu*[1,2], *Li Yang*[1,2], *Wen Jin*[1,2], *Bichen Xiao*[1,2], *Wenfeng Zhang*[1,2,3], *Haixin Chang*[1,2,3,*]

[1]State Key Laboratory of Material Processing and Die & Mold Technology, School of Materials Science and Engineering, Huazhong University of Science and Technology, Wuhan 430074, China.

[2]Wuhan National High Magnetic Field Center and Institute for Quantum Science and Engineering, Huazhong University of Science and Technology, Wuhan 430074, China.

[3]Shenzhen R&D Center of Huazhong University of Science and Technology, Shenzhen 518000, China.

[*]Corresponding authors. E-mail: hxchang@hust.edu.cn

**Room-temperature electrically-tuned coercivity and nonvolatile multi-states magnetization switching is crucial for next-generation low-power 2D spintronics. However, most methods have limited ability to adjust the coercivity of ferromagnetic systems, and room-temperature electrically-driven magnetization switching shows high critical current density and high power dissipation. Here, highly-tunable coercivity and highly-efficient nonvolatile multi-states magnetization switching are achieved at room temperature in single-material based devices by 2D van der Waals itinerant ferromagnet $Fe_3GaTe_2$. The coercivity can be readily tuned up to ~98.06% at 300 K by a tiny in-plane electric field that is 2-5 orders of magnitude smaller than that of other ferromagnetic systems. Moreover, the critical current density and power dissipation for room-temperature magnetization switching in 2D $Fe_3GaTe_2$ are down to ~$1.7\times10^5$ $A\cdot cm^{-2}$ and ~$4\times10^{12}$ $W\cdot m^{-3}$, respectively. Such switching power dissipation is 2-6 orders of magnitude lower than that of other 2D ferromagnetic systems. Meanwhile, multi-states magnetization switching are presented by continuously controlling the current, which can dramatically enhance the information storage capacity and develop new computing methodology. This work opens the avenue for room-temperature electrical control of ferromagnetism and potential applications for vdW-integrated 2D spintronics.**

## 1. Introduction

Room-temperature electrical modulation of magnetism and magnetization switching are technically important for practical applications of spintronics[1-7]. The magnetization switching are usually performed through a current-induced spin-torque, which has been used in some multifunctional spintronic applications[8-10]. However, most electrical methods have limited ability to adjust the coercivity ($H_C$) of ferromagnetic systems[6, 11-13], and electrically-driven magnetization switching exhibits high critical current density and high power dissipation at room temperature[3, 14-17]. Since 2017, 2D van der Waals (vdW) ferromagnetic crystals have attracted colossal attention in the field of fundamental physics and spintronics[18-23]. Electrical modulation of magnetism in 2D vdW magnetic crystals is essential for next-generation vdW-integrated spintronics[11, 12, 24, 25]. Moreover, the current-tunable coercivity through spin-orbit torques (SOT) makes it possible to achieve high efficient nonvolatile magnetization switching with multi-level states in spintronic devices, which has been demonstrated in single ferromagnet $Fe_3GeTe_2$[26-28]. But the Curie temperature ($T_C$) of $Fe_3GeTe_2$ is much below room temperature, which limits its practical application. In contrast, 2D vdW ferromagnetic crystals $Fe_3GaTe_2$ shows above-room-temperature $T_C$, large perpendicular magnetic anisotropy and potential spintronic application prospects[29-33], which is expected to promote this high efficiency and multi-states nonvolatile magnetization switching up to room temperature.

Here, by using 2D vdW room-temperature ferromagnetic $Fe_3GaTe_2$ crystals with itinerant ferromagnetism and intrinsic anomalous Hall effect (AHE), we report highly-tunable coercivity and highly-efficient nonvolatile multi-states magnetization switching achieved by small current at room temperature. The coercivity of 2D $Fe_3GaTe_2$ can be easily regulated ~98.06% at 300 K by applying a very small in-plane electric field of ~0.029 $mV \cdot nm^{-1}$, which is 2-5 orders of magnitude smaller than that of other ferromagnetic systems. Moreover, the nonvolatile magnetization switching by current is realized at room temperature with high efficiency. The room-temperature switching critical current density is lower than that of all known spin-torques induced magnetization switching systems, and the power dissipation is 2-6 orders of magnitude lower than that of other 2D ferromagnetic systems. Importantly, multi-states magnetization switching are realized by continuously tuning current, which is important for high-density data storage and high-efficiency computing based on 2D vdW ferromagnetic crystals.

## 2. Results

### 2.1. Structure, itinerant ferromagnetism and intrinsic AHE

High-quality $Fe_3GaTe_2$ single crystals are grown by self-flux method and adopted for the device fabrication (see **Experimental Section**). The as-grown $Fe_3GaTe_2$ crystals feature vdW-stacked structure along the *c* axis and space group P63/mmc (**Figure 1a**)[29]. As shown in **Figure S1a** (**Supporting Information**), the as-grown $Fe_3GaTe_2$ crystals are millimetre-sized, shiny, highly crystalline with oriented growth along (0 0 2L) planes, confirmed by optical microscopy and X-ray diffraction (XRD) pattern. Moreover, **Figure S1b** (**Supporting Information**) shows the high-resolution transmission electron microscopy (HRTEM) with clear crystal lattices and corresponded fast Fourier transform (FFT) images taken along the [001] zone axis, matching well with the hexagonal structure and single crystal nature of $Fe_3GaTe_2$. The interlayer distance of (100) plane is measured as 0.346 nm.

As illustrated in **Figure S2a** (**Supporting Information**), the as-grown vdW $Fe_3GaTe_2$ ferromagnetic crystals exhibit typical ferromagnetic-paramagnetic transition at above room temperature. Temperature-dependent isothermal magnetization (M-B) curves with a series of hysteresis loops also confirm this result (**Figure S2b, Supporting Information**), similar to the previous report[29]. The saturation magnetization ($M_S$) at 2 K is calculated as $28.3 \times 10^3$ emu · mol$^{-1}$ (1.68 μB/Fe). Moreover, the itinerant ferromagnetism of the $Fe_3GaTe_2$ crystal is identified by Takahashi theory of spin fluctuations[34] and self-consistent renormalization (SCR) theory[35] with the aid of $M^4$-B/M plot (**Figure S2c**, more discussions in **Note S1, Supporting Information**). Thus, single-crystal $Fe_3GaTe_2$ is identified as an intrinsic, above-room-temperature, itinerant 2D ferromagnet with large PMA, which is suitable for effectively electrically-tuned ferromagnetism due to the shielding effect of localized magnetic moments by complex interplay between itinerant electrons and localized atomic magnetic moments[36].

In this work, four 2D $Fe_3GaTe_2$-based Hall devices are used for subsequently temperature- and current-tuned magneto-transport tests, and their thickness are identified by atomic force microscopy (AFM) (**Figure S3, Supporting Information**). The optical image of a typical 10 nm $Fe_3GaTe_2$ Hall device is presented in **Figure 2b**. Temperature-dependent longitudinal resistance ($R_{xx}$-T) curves of 10 and 27.2 nm $Fe_3GaTe_2$ nanosheets show the typical metallic feature and ferromagnetic-paramagnetic secondary-phase transition, identifying the corresponded $T_C$ of ~340 and ~360 K, respectively (**Figure 2c**). Moreover, clear AHE from 2 to over 300 K are presented by magnetic field-dependent Hall resistance ($R_{xy}$-B) curves under different temperatures (**Figure 2d, e**). The nearly square hysteresis loops indicates that large PMA is well preserved for $Fe_3GaTe_2$ with thickness down to 10 nm. Consistent with the $R_{xx}$-T results, the $T_C$ decreases from ~360 to ~340 K when the thickness of 2D $Fe_3GaTe_2$ nanosheets goes down from 27.2 to 10 nm (**Figure 2f**), implying the typical influence of thermal fluctuation on 2D ferromagnetic order[37]. Further, at temperatures below 100 K, the anomalous Hall conductivity ($\sigma_{AH}$) of 27.2 and 10 nm $Fe_3GaTe_2$ nanosheets is revealed to be independent of both temperature and conductivity ($\sigma_{xx}$) (**Figure 2g**). This robustness suggests that the observed AHE in 2D $Fe_3GaTe_2$ is not governed by scattering events, but arises from intrinsic mechanism and is therefore related to Berry curvature in momentum space, as expected in a unified model framework of AHE physics[38-41].

## 2.2. Room-temperature highly-tunable $H_C$ by current

To study the current-induced effects in 2D $Fe_3GaTe_2$ nanosheets, AHE under different pulse currents have been carefully measured by standard testing geometry (**Figure 2a**). At room temperature, the $H_C$ of the hysteresis loop gets significant reduction as an in-plane current applied to the 10 nm $Fe_3GaTe_2$ from 0.005 to 2.5 mA (or -0.005 to -2.5 mA) (**Figure 2b**). Note that increasing the current to ±2.5 mA significantly reduces the $H_C$ with the almost unchanged $R_{AH}$ (**Figure 2c**), suggesting that the $M_S$ of this 10 nm $Fe_3GaTe_2$ is not much affected by applied current. Also, the hysteresis loop is respectively measured 10 times under current of 0.5, 1, 1.5 mA and the similar results are obtained, showing the good repeatability of such current-induced effect without attenuation (**Figure S4, Supporting Information**). Thus, the significant reduction of $H_C$ by current at room temperature do not disrupt the ferromagnetic order of $Fe_3GaTe_2$, which is desirable for 2D spintronic applications.

Similar phenomenon of current-controlled $H_C$ with almost unchanged $R_{AH}$ has also been studied in 2D $Fe_3GaTe_2$ nanosheets under different thicknesses and temperatures (**Figure S5 and S6, Supporting Information**). At room temperature, with the increase of thickness of 2D $Fe_3GaTe_2$ from 10 to 27.2 nm, the regulated percentage of $H_C$ is gradually decrease (**Figure 2d**). For the 10 nm $Fe_3GaTe_2$, with the increase of temperature from 2 to 300 K, the regulated percentage of $H_C$ firstly decreases and then increases gradually (**Figure 2e**). Based on these, the largest regulated percentage of $H_C$ is achieved in 10 nm $Fe_3GaTe_2$ at 300 K, which is up to ~98.06% by applying a current of 2.5 mA (≈0.029 $mV{\cdot}nm^{-1}$ in-plane electric field), showing unprecedented advantages compared with all known electrically-controlled ferromagnetism systems (**Figure 2f**). For example, all previous reports about electrically-controlled $H_C$ by electrostatic or ferroelectric gating can only achieve a low regulated percentage of $H_C$ (≤61.4%) through a very large electric field (~1-$10^3$ mV·$nm^{-1}$)[6, 11-13], which is 2-5 orders of magnitude larger than required electric field of current-controlled $H_C$ in this work (**Table S1, Supporting Information**). Therefore, such results of the efficient and low-power modulation of the $H_C$ in 2D $Fe_3GaTe_2$ provides a promising alternative for room-temperature electrical control of ferromagnetism.

**2.3. Highly-efficient nonvolatile multi-states magnetization switching at room temperature**

The current-induced $H_C$ reduction can arise from the Joule heating effect since $H_C$ reduces at higher temperatures. However, this possibility has been carefully evaluated and eliminated (**Figure S7**, more discussions in **Note S2**, **Supporting Information**). Then, based on room-temperature ferromagnetic 2D $Fe_3GaTe_2$ nanosheet (10 nm) with current-induced $H_C$ reduction, we design a type of magnetic memory device, where data information can be written by current and read through the $R_{xy}$ with the aid of a tiny magnetic field (**Figure 3a**). The positive and negative value of saturated $R_{xy}$ are defined as "1" ("+$M_S$" state) and "0" ("-$M_S$" state) states, respectively, and the $\Delta R_{xy}$ is ~7.8 Ω. Specifically, in a tiny positive magnetic field, magnetization changes from the "0" state to the "1" state along path I by increasing the current. In a tiny negative magnetic field, magnetization returns from the "1" state to the "0" state along path II by increasing the current. **Figure 3b** exhibits the corresponded room-temperature magnetization switching as a function of applying current. Under +40 Oe, when current increases from 0.005 to 2.5 mA, the magnetization gradually increases from “0” state to “1” state (represented by the red arrow). It is worth noting that after reaching the “1” state, the magnetization remains unchanged no matter how the current is swept, which indicates nonvolatile behavior when the magnetic information is transformed from the "0" state to the "1" state via current. Only by switching the magnetic field to -40 Oe can the magnetization be controlled from the "1" state return to the "0" state with the increase of the current (represented by the blue arrow). Moreover, this nonvolatile magnetization switching behavior still exists at temperatures below room temperature (2-250 K) (**Figure S8, Supporting Information**).

In order to study the influence of the assistant magnetic field on the magnetization switching, we perform the magnetization switching under different assistant magnetic fields (|B|=0-50 Oe) by using a 10 nm $Fe_3GaTe_2$ nanosheet at room temperature, as shown in **Figure 3c**. The behavior of magnetization switching are observed in all applied magnetic field, while only when the magnetic field |B|≥40 Oe can achieve magnetization switching close to 100% (~96%). It is worth noting that such magnetization switching close to 100% at 300 K has no attenuation after 100 cycles, indicating the robust property of current-driven nonvolatile magnetization switching in 2D $Fe_3GaTe_2$ (**Figure S9, Supporting Information**). As shown in **Figure 3d**, with the increase of the assistant magnetic field, the switching ratio gradually increases and the critical current density ($J_C$) for magnetization switching gradually decreases, and eventually both reach saturation. In addition, the current-driven nonvolatile magnetization switching is also performed at room temperature on other three $Fe_3GaTe_2$ nanosheets with different thickness (**Figure S10, Supporting Information**), and their corresponded $J_C$ are presented in **Figure 3e**.

Furthermore, the intermediate multi-level magnetization states between “0” and “1” states is realized at room temperature by simply controlling the writing current in the 10 nm $Fe_3GaTe_2$ nanosheet (**Figure 3f, g**). To be specific, roughly 8 and 9 step-like resistance states are presented by continuously controlling the writing current from 0.005 to 2 mA under -40 Oe and +40 Oe magnetic field, respectively. The corresponding $R_{xy}$ values as a function of writing current are summarized in **Figure 3h**. Please note that the possible influence of Barkhausen effect on the multi-level states in 2D $Fe_3GaTe_2$ device is negligible because each intermediate state is stable for a very long time (≥170 s) under constant current[27]. Therefore, this continuous electrically-tunable multi-states property in 2D $Fe_3GaTe_2$ can serve as a new type of multiple-state spin memory. Such multiple states enable several digitals (multi bits) more than conventional two digitals (1 bit), which not only highly enhances the information storage density, but also improve the computing efficiency.

For a memory device, the critical current density $J_C$ and the switching power dissipation $J_C^2/\sigma$ ($\sigma$ is conductivity) are two important indicators used for energy efficiency[27].

**Figure 4a** summarizes the temperature-dependent $J_C$ of magnetization switching in various 2D ferromagnetic systems. 2D $Fe_3GaTe_2$ shows lowest $J_C$ (~$1.7\times10^5$ $A\cdot cm^{-2}$) at room temperature, lower than that of all known room-temperature spin-torques induced magnetization switching systems[2, 5, 42, 43]. Meanwhile, the lowest switching power dissipation $J_C^2/\sigma$ of 2D $Fe_3GaTe_2$ are on the order of ~$4\times10^{12}$ $W\cdot m^{-3}$, 2-6 orders of magnitude lower than that of other 2D room-temperature ferromagnetic systems[3, 14-17] (**Figure 4b**). These performance highlights the high energy efficiency of our 2D $Fe_3GaTe_2$ devices at room temperature, which is long-sought for practical applications of 2D spintronics.

## 3. Conclusions

In summary, highly-tunable coercivity and highly-efficient nonvolatile multi-states magnetization switching has been realized at room temperature in single 2D vdW itinerant ferromagnet $Fe_3GaTe_2$. The 2D $Fe_3GaTe_2$ nanosheets show very large regulated percentage of $H_C$ of ~98.06% under a tiny in-plane electric field that is 2-5 orders of magnitude smaller than required electric field of other ferromagnetic systems. In addition, the room-temperature nonvolatile magnetization switching are achieved at a small magnetic field (~40 Oe) with a low $J_C$ of ~$1.7\times10^5$ $A\cdot cm^{-2}$ and an ultralow power dissipation of ~$4\times10^{12}$ $W\cdot m^{-3}$, far superior to that of all known 2D ferromagnetic systems. Multi-states magnetization switch has also been demonstrated at room temperature by simply controlling the current in the 10 nm $Fe_3GaTe_2$ nanosheet. These findings open up a promising avenue for room-temperature electrically-controlled magnetism and spintronic applications based on 2D vdW ferromagnetic crystals.

## 4. Experimental Section

*Device fabrication:* High-quality $Fe_3GaTe_2$ single crystals were grown by self-flux method[29]. To fabricate the device, Cr/Au (8/17 nm) electrodes were firstly prepared on a $SiO_2$/Si substrate by laser direct writing system (MicroWriter ML3, DMO), electron beam evaporation and lift-off process. Subsequently, the exfoliated 2D $Fe_3GaTe_2$ nanosheets were transferred onto the pre-fabricated electrodes using a homemade micro-transfer system. Entire exfoliation and transfer process were conducted in an Ar-filled glove box ($H_2O$, $O_2$ < 0.1 ppm) to avoid oxidation.

*Characterizations and measurements:* The morphology, structure, and thickness of $Fe_3GaTe_2$ were performed at room temperature by Optical microscopy (OM, MV6100), powder X-ray diffraction (XRD, Smartlab SE, Rigaku Corporation), field-emission transmission electron microscopy (FTEM, Tecnai G2 F30, FEI) and atomic force microscopy (AFM, Dimension EDGE, Bruker). The magnetic properties and magneto-transport of $Fe_3GaTe_2$ crystals were measured using a physical property measurement system (PPMS-14 T, Quantum Design, USA) with the out-of-plane magnetic field unless otherwise stated. Each resistance data was tested 25 times for an average with a small constant reading current $I_{read}$=0.005-0.02 mA depending on the thickness of samples. For current-tunable test, a millisecond writing current pulse was applied, and then, $I_{read}$ was applied to read out the resistance after waiting for several seconds.

## Supporting Information

Supporting Information is available from the Wiley Online Library or from the author.

## Acknowledgements

H.C. designed the project. G.Z. and W.J. grew the single crystals. G.Z. fabricated the Hall devices. G.Z., H.W. and L.Y. did the characterization and measurements. H.C., G.Z. and W.Z. analyzed the results. G.Z. and H.C. wrote the paper. This work was supported by the National Key Research and Development Program of China (2022YFE0134600), National Natural Science Foundation of China (52272152, 61674063 and 62074061), Shenzhen Science and Technology Innovation Committee (JCYJ20210324142010030), Natural Science Foundation of Hubei Province, China (2022CFA031), and Interdisciplinary Research Program of Huazhong University of Science and Technology (5003110122).

## Conflict of Interest

The authors declare no conflict of interest.

## Data Availability Statement

The data that support the findings of this study are available from the corresponding author upon reasonable request.

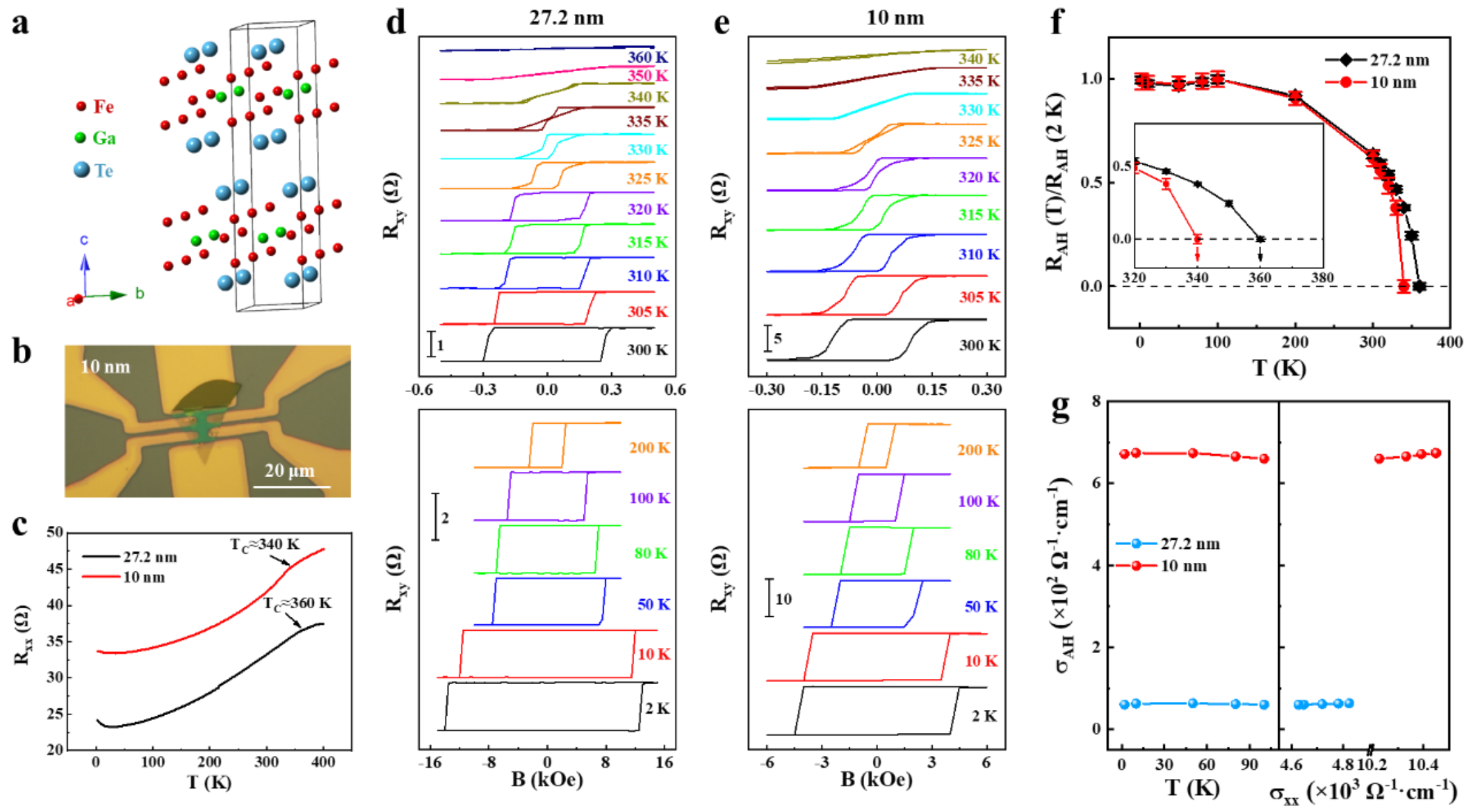

**Figure 1. Structure, magneto-transport and intrinsic AHE of 2D vdW room-temperature ferromagnetic $Fe_3GaTe_2$ crystals.** (a) Crystal structure of vdW $Fe_3GaTe_2$ crystal. (b) Optical image of the 10 nm $Fe_3GaTe_2$ Hall device. (c) Temperature-dependent longitudinal resistance ($R_{xx}$-T) curves of 27.2 and 10 nm $Fe_3GaTe_2$. (d, e) Magnetic field- dependent Hall resistance ($R_{xy}$-B) curves under different temperatures of 27.2 and 10 nm $Fe_3GaTe_2$. (f) Temperature-dependent normalized anomalous Hall resistance ($R_{AH}(T)/R_{AH}(2\ K)$). Inset is the enlarge image around $T_C$. Error bars SD, N=25. (g) Temperature- and conductivity ($\sigma_{xx}$)-dependent anomalous Hall conductivity ($\sigma_{AH}$) below 100 K. The $\sigma_{AH}=\rho_{AH}/((\rho_{AH})^2+(\rho_{xx})^2)$ and $\sigma_{xx}=1/\rho_{xx}$, where $\rho_{AH}$ and $\rho_{xx}$ are anomalous Hall resistivity and longitudinal resistivity, respectively.

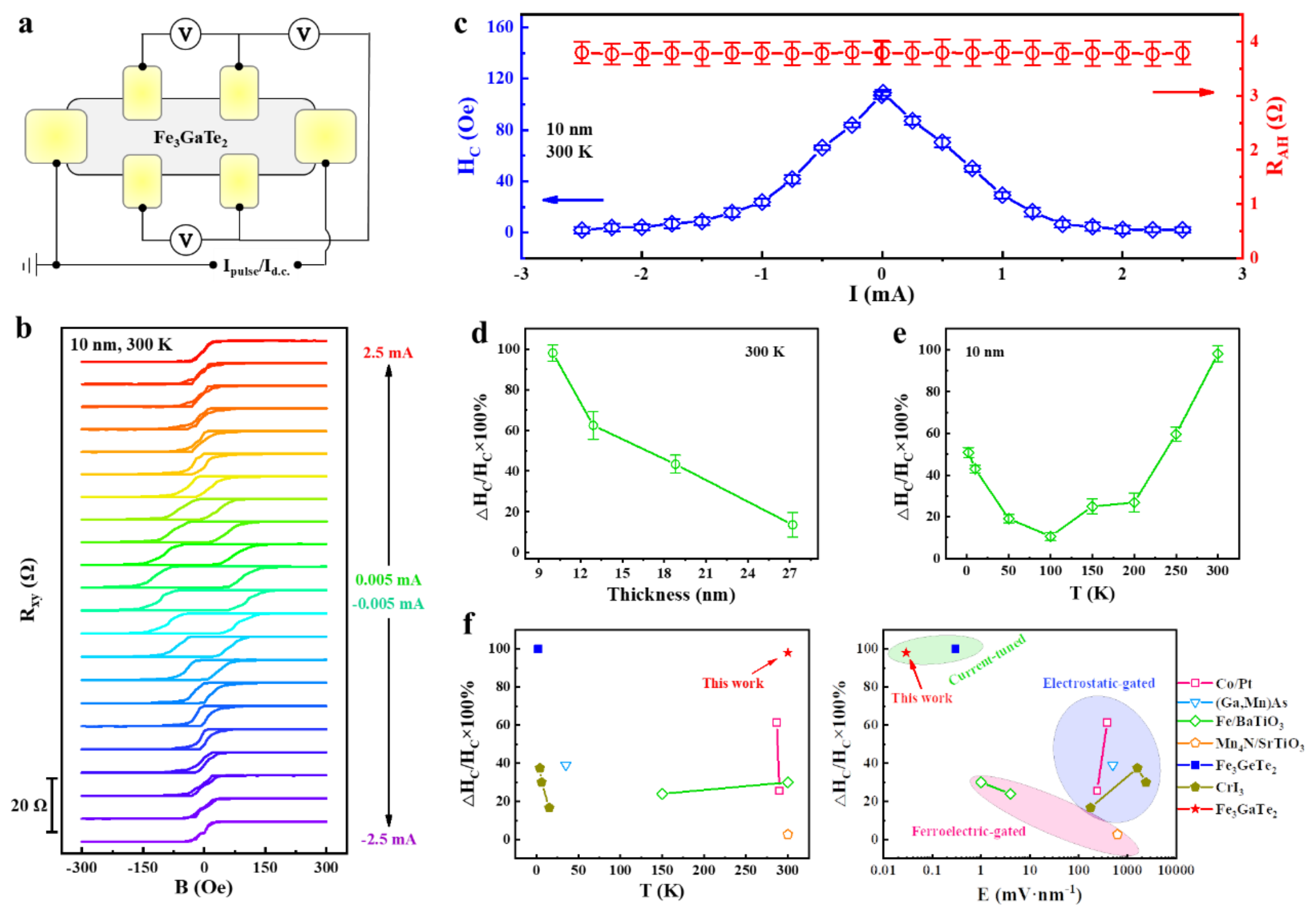

**Figure 2. Room-temperature highly-tunable $H_C$ by current in 2D $Fe_3GaTe_2$ nanosheets.** (a) Schematic and measurement geometry of the 2D $Fe_3GaTe_2$-based Hall device. (b) $R_{xy}$-B curves under different currents measured at 300 K in 10 nm $Fe_3GaTe_2$. (c) Current-dependent $H_C$ and $R_{AH}$ at 300 K in 10 nm $Fe_3GaTe_2$. Error bars SD, N=3 for $H_C$ and N=25 for $R_{AH}$. (d, e) Thickness- and temperature-dependent regulated percentage of $H_C$ ($\Delta H_C/H_C \times 100\%$) by current in 2D $Fe_3GaTe_2$ nanosheets. The $\Delta H_C/H_C$=[$H_C$(2.5 mA)-$H_C$(smallest I)]/$H_C$(smallest I). Error bars SD, N=3. (f) Comparison of working temperature, regulated percentage of $H_C$ and required electric field with various ferromagnetic systems. The solid and open symbols represent the vdW and non-vdW magnets, respectively. See data and references in **Table S1, Supporting Information**.

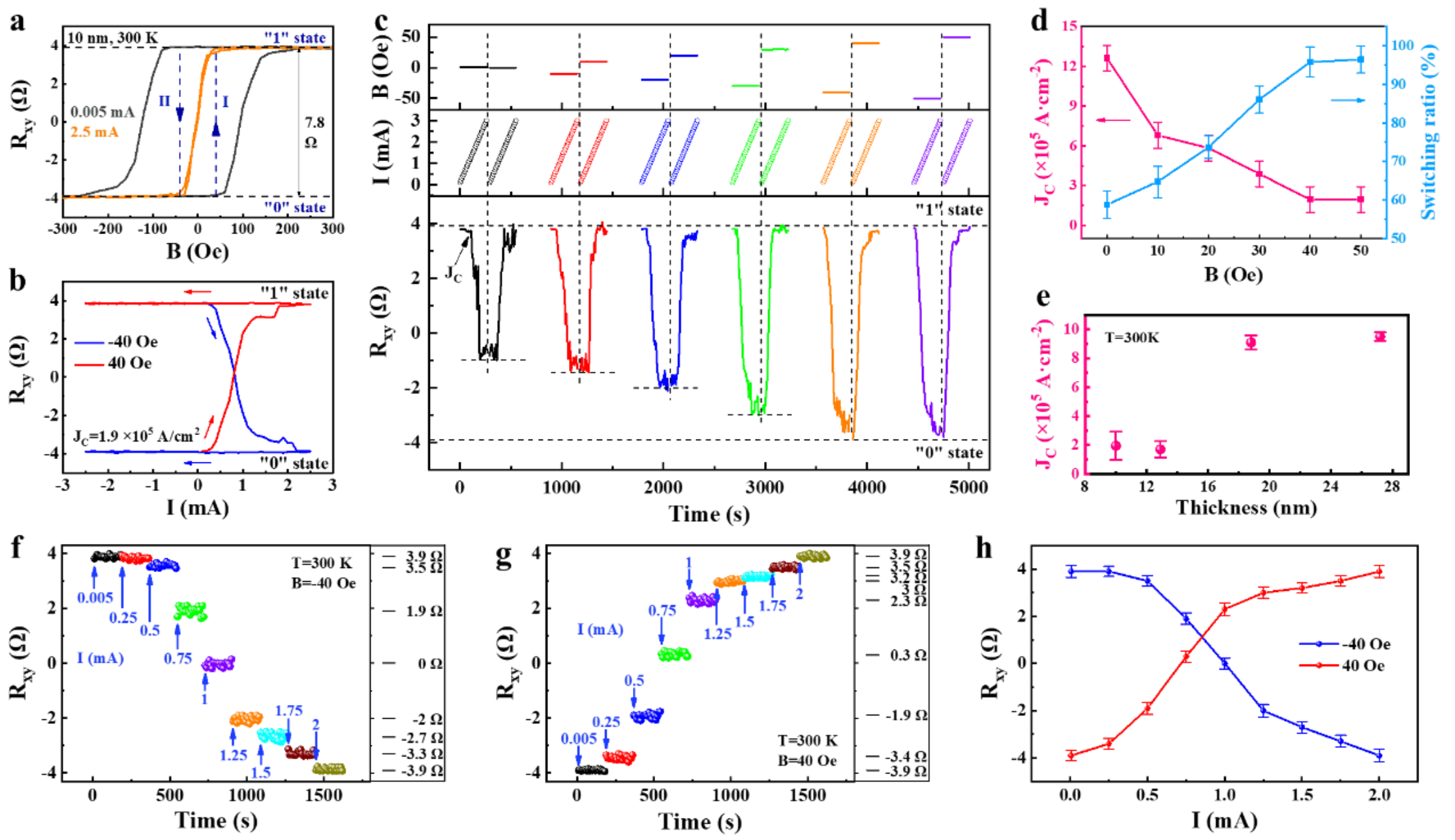

**Figure 3. Room-temperature ultralow-current driven nonvolatile magnetization switching with multilevel states in 2D $Fe_3GaTe_2$ nanosheets.** (a) $R_{xy}$-B curves of 10 nm $Fe_3GaTe_2$ with applied current I=0.005 and 2.5 mA. The arrows illustrate the transition between “0” and “1” states by the writing current through a novel-writing path. (b) Ultralow-current driven nonvolatile magnetization switching between “0” and “1” states of 10 nm $Fe_3GaTe_2$ assisted by tiny magnetic field ±40 Oe. (c) Current-driven magnetization switching of 10 nm $Fe_3GaTe_2$ under different assistant tiny magnetic field from 0 to ±50 Oe. (d) $J_C$ and switching ratio of 10 nm $Fe_3GaTe_2$ under different assistant magnetic field. (e) $Fe_3GaTe_2$ thickness-dependent $J_C$. (f, g) Time-dependent $R_{xy}$ with different current under -40 and +40 Oe, respectively. The blue arrows denote the value of applied current as time passed. (h) Multi-level $R_{xy}$ under each current. Error bars SD, N=25. All data are measured at room temperature (300 K). The error bars in d and e are obtained from the step of applied current (0.1 mA).

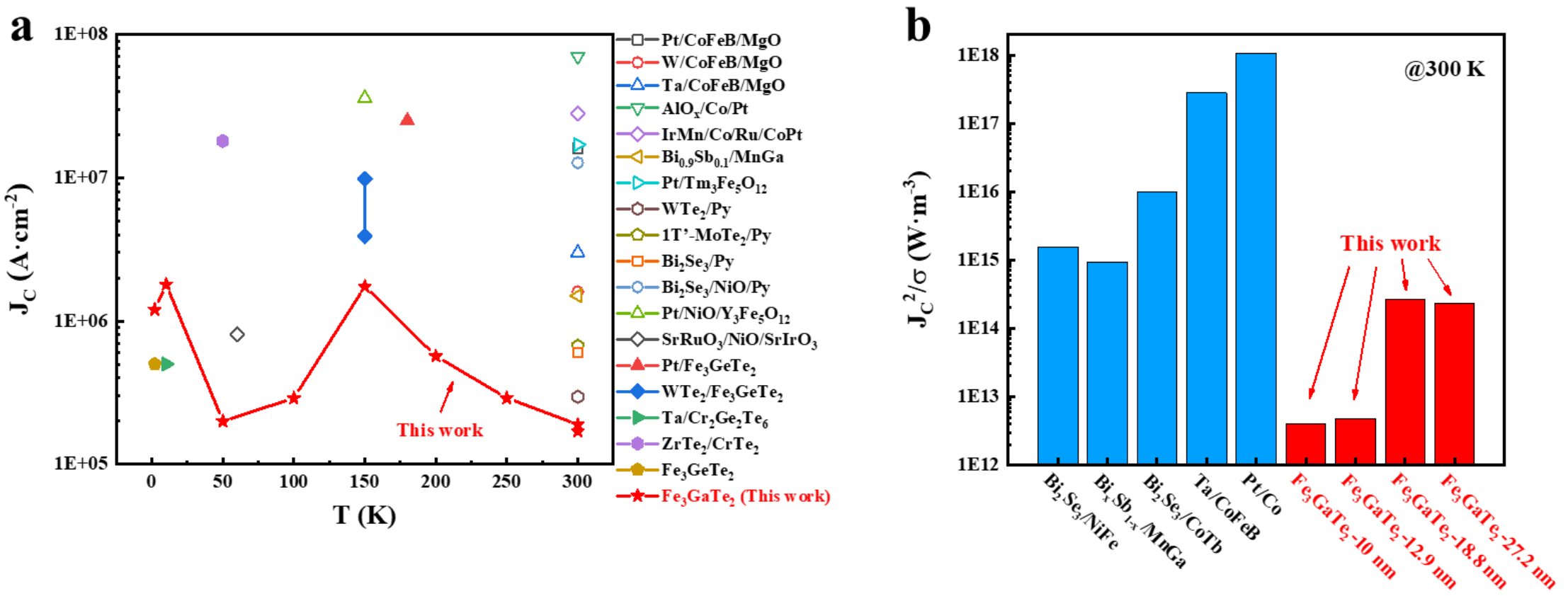

**Figure 4. Comparison of magnetization switching with that of other 2D ferromagnetic systems.** (a) Working temperature and $J_C$ of 2D $Fe_3GaTe_2$ and other 2D ferromagnetic systems. The solid and open symbols represent the devices based on vdW and non-vdW ferromagnets, respectively. See data and references in **Table S2, Supporting Information**. (b) Switching power dissipation $J_C^2/\sigma$ of 2D $Fe_3GaTe_2$ and other 2D ferromagnetic systems at room temperature[3, 14-17].

# Supporting Information for

# Room-temperature Highly-Tunable Coercivity and Highly-Efficient Nonvolatile Multi-States Magnetization Switching by Small Current in Single 2D Ferromagnet $Fe_3GaTe_2$

*Gaojie Zhang, Hao Wu, Li Yang, Wen Jin, Bichen Xiao, Wenfeng Zhang, Haixin Chang*,*

*Corresponding author. E-mail: hxchang@hust.edu.cn

**This file includes:**

Notes S1-S2

Figures S1-S10

Tables S1-S2

**Notes S1-S2:**

**Note S1.** Identification of itinerant ferromagnetism in $Fe_3GaTe_2$

**Note S2.** Evaluation of Joule heating effect

**Figures S1-S10:**

**Figure S1.** Structural characterizations of vdW $Fe_3GaTe_2$ crystals

**Figure S2.** Above-room-temperature strong ferromagnetism in vdW $Fe_3GaTe_2$ bulk crystals by VSM

**Figure S3.** AFM characterizations of four as-tested 2D $Fe_3GaTe_2$ Hall devices

**Figure S4.** Repeatability of room-temperature current-controlled $H_C$ under different applied current in the 10 nm $Fe_3GaTe_2$ nanosheet

**Figure S5.** Optical images and room-temperature current-controlled $H_C$ in $Fe_3GaTe_2$ nanosheets with different thickness

**Figure S6.** Current-controlled $H_C$ in 10 nm $Fe_3GaTe_2$ nanosheet under different temperatures

**Figure S7.** Elimination of Joule heating effect

**Figure S8.** Current-driven nonvolatile magnetization switching in the 10 nm $Fe_3GaTe_2$ nanosheet under different temperatures

**Figure S9.** Cyclic stability of ≈100% magnetization switching of 10 nm $Fe_3GaTe_2$ after 100 times cycles at room temperature

**Figure S10.** Room-temperature current-driven nonvolatile magnetization switching in 2D $Fe_3GaTe_2$ nanosheets with different thickness

**Tables S1-S2:**

**Table S1.** Comparison of working temperature, required electric field, and tunable ratio of $H_C$ with various ferromagnetic systems

**Table S2.** Comparison of working temperature and $J_C$ of magnetization switching with various ferromagnetic systems

## Note S1. Identification of itinerant ferromagnetism in $Fe_3GaTe_2$

The $M^4$-B/M are plotted to determine whether the $Fe_3GaTe_2$ belongs to the itinerant or a localized spin system (**Figure S2c**). According to the Takahashi theory of spin fluctuations, the relationship between $M^4$ and B/M can be described as[1]:

$$M^4=1.17\times10^{18}\left(\frac{T_C^2}{T_A^3}\right)\left(\frac{B}{M}\right) \tag{1}$$

where M is the magnetization, $T_A$ is the dispersion of the spin fluctuation spectrum in wave vector space. The linear slope estimated from the $M^4$-B/M plot determines the $T_C$=348 K and $T_A$=195 K for $Fe_3GaTe_2$. According to the self-consistent renormalization (SCR) theory, $T_C$ can be expressed as[2]:

$$T_C=(60c)^{-3/4}M_S^{3/2}T_A^{3/4}T_0^{1/4} \tag{2}$$

where c≈0.3353, $M_S$ is the saturation magnetization in μB/Fe unit at low temperature, and $T_0$ is the dispersion of the spin fluctuation spectrum in energy space. Thus, we obtained the $T_0=7.16\times10^5$ K for $Fe_3GaTe_2$. The ratio of $T_C/T_0$, a crucial parameter characterizing the localized ($T_C/T_0\approx1$) or itinerant ($T_C/T_0 \ll 1$) character of ferromagnetism[3], is calculated as $\sim4.9\times10^{-4}$ in $Fe_3GaTe_2$. Therefore, we identify the $Fe_3GaTe_2$ is an above-room-temperature vdW itinerant ferromagnet.

## Note S2. Evaluation of Joule heating effect

The Joule heating effect is evaluated by $R_{xx}$ and further verified by $R_{xy}$-B curves at different temperatures and currents. Here, the $R_{xx}$ under each current is extracted from the $R_{xx}$-B curves at 300 K and zero magnetic field. Since the $Fe_3GaTe_2$ is a ferromagnetic metal, a gradual increase in temperature from 300 K will increase the $R_{xx}$, as shown in **main text Figure 1c**. Therefore, if the Joule heating effect is significant in $Fe_3GaTe_2$, the $R_{xx}$ should increase with the increase of current. However, as shown in **Figure S7a**, no obvious increase of $R_{xx}$ is observed as increasing current for four as-tested $Fe_3GaTe_2$ nanosheets, indicating the negligible Joule heating effect.

To further verify the negligible Joule heating effect, we compare the changes of hysteresis loops caused by increasing temperature and current respectively. As shown in two upper panels in **Figure S7b, c**, raising temperature from 300 K will reduce both $H_C$ and saturated $R_{xy}$ of 2D $Fe_3GaTe_2$. Meanwhile, as shown in lower panels in **Figure S7b, c**, we can also reduce $H_C$ at 300 K by increasing the current and achieve the similar effect as raising temperature. Interestingly, however, the increase of current did not simultaneously reduce the saturated $R_{xy}$. These results further confirm that the Joule heating effect in this work is negligible and suggest that the regulating effects of current and temperature on hysteresis loops belong to two totally different mechanisms. Similar evaluation methods have also been reported in other literature[4].

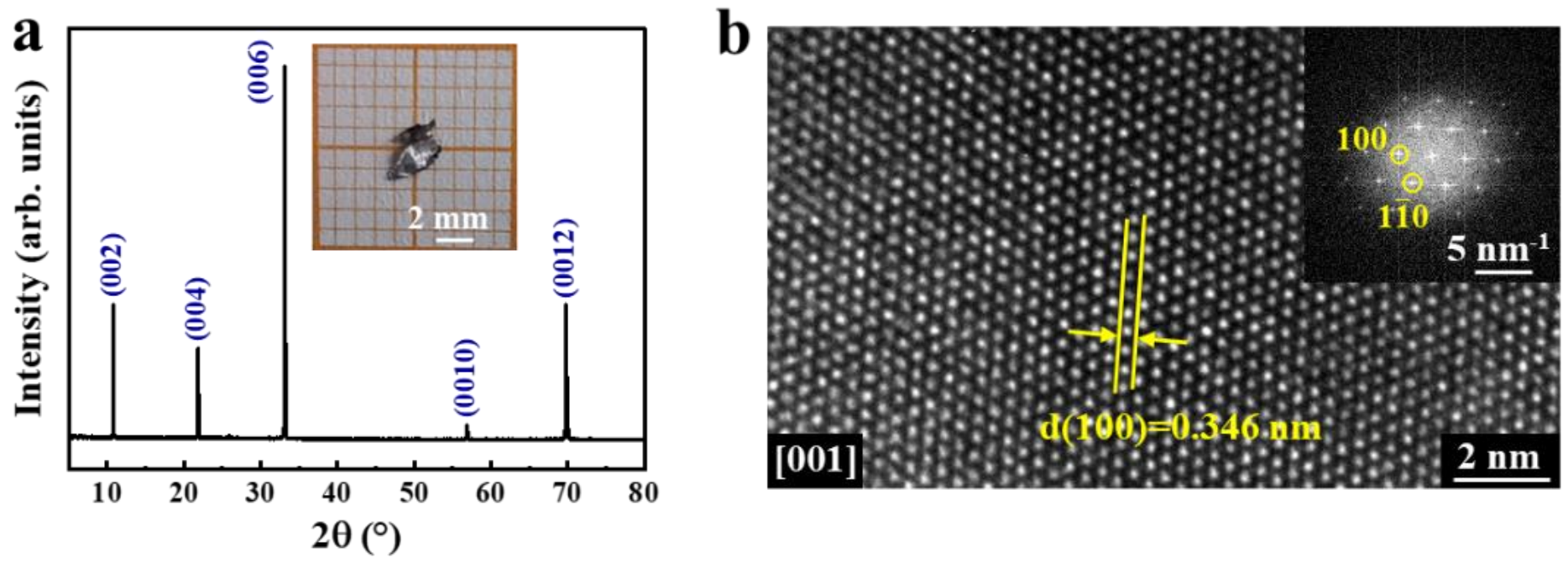

**Figure S1. Structural characterizations of vdW $Fe_3GaTe_2$ crystals. a,** XRD pattern of bulk $Fe_3GaTe_2$ crystals. Inset shows a typical crystal photograph. **b**, HRTEM image and corresponded FFT pattern of the $Fe_3GaTe_2$ single crystal.

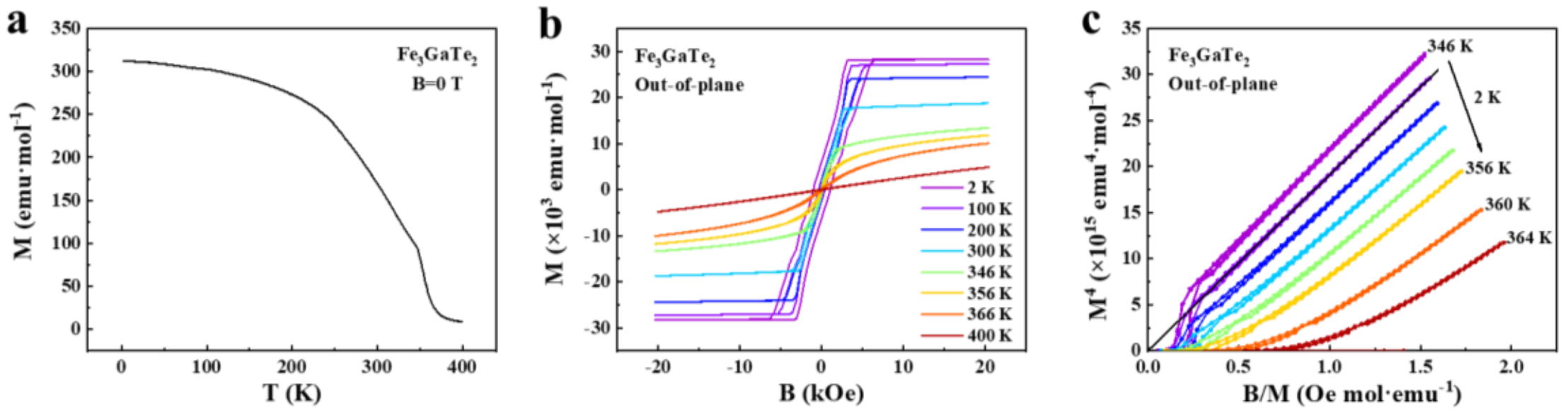

**Figure S2. Above-room-temperature strong ferromagnetism in vdW $Fe_3GaTe_2$ bulk crystals by VSM. a**, Temperature-dependent magnetization (M-T) curve of $Fe_3GaTe_2$ crystals without external magnetic field. **b**, Isothermal magnetization (M-B) curves at varying temperatures from 2 to 400 K with out-of-plane magnetic field. **c**, $M^4$ vs B/M plots for $Fe_3GaTe_2$ crystals at various temperatures around $T_C$. The linear fitting curve is used to analyze the Takahashi theory of spin fluctuations.

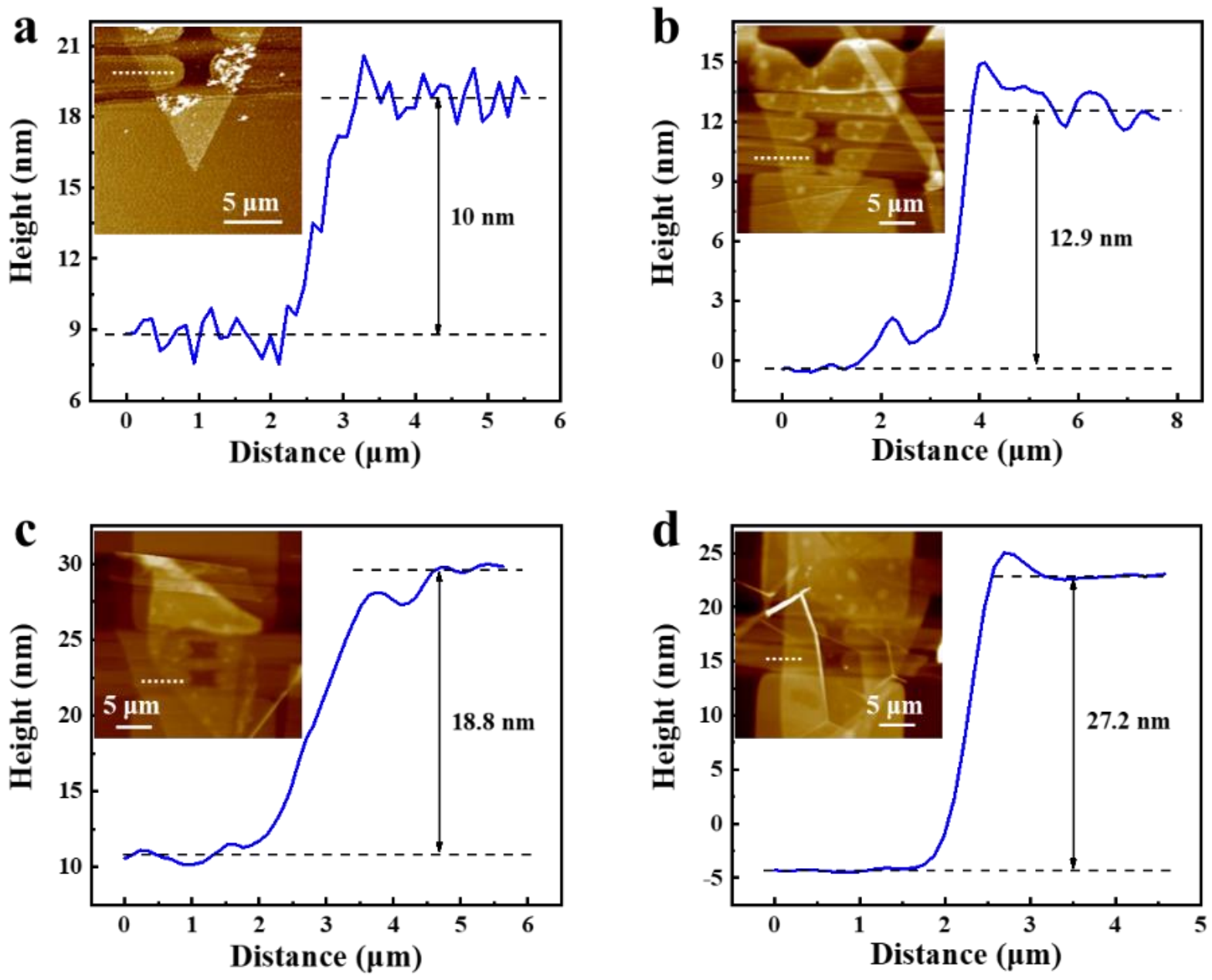

**Figure S3. AFM characterizations of four as-tested 2D $Fe_3GaTe_2$ Hall devices. a**, 10 nm. **b**, 12.9 nm. **c**, 18.8 nm. **d**, 27.2 nm.

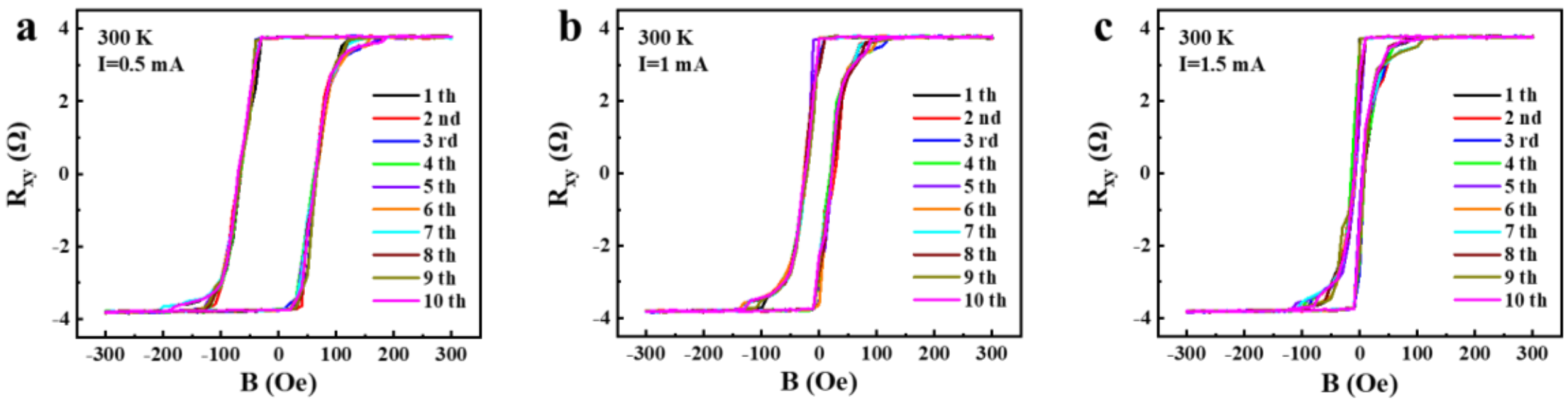

**Figure S4. Repeatability of room-temperature current-controlled $H_C$ under different applied current in the 10 nm $Fe_3GaTe_2$ nanosheet. a**, 0.5 mA. **b**, 1 mA. **c**, 1.5 mA. Each hysteresis loop is tested 10 times to ensure the repeatability.

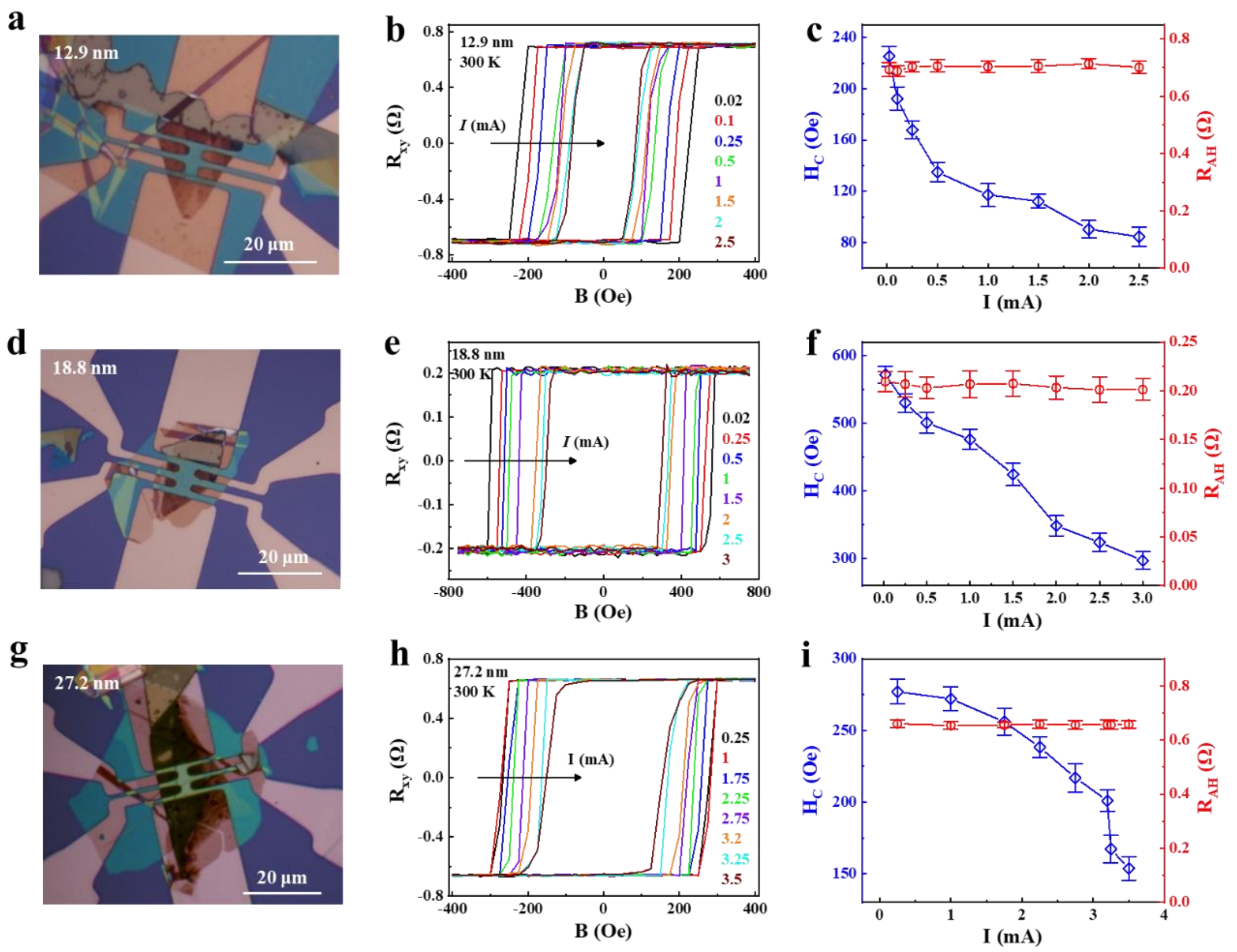

**Figure S5. Optical images and room-temperature current-controlled $H_C$ in $Fe_3GaTe_2$ nanosheets with different thickness. a-c**, 12.9 nm. **d-f**, 18.8 nm. **g-i**, 27.2 nm. Error bars SD, N=3 for $H_C$ and N=25 for $R_{AH}$.

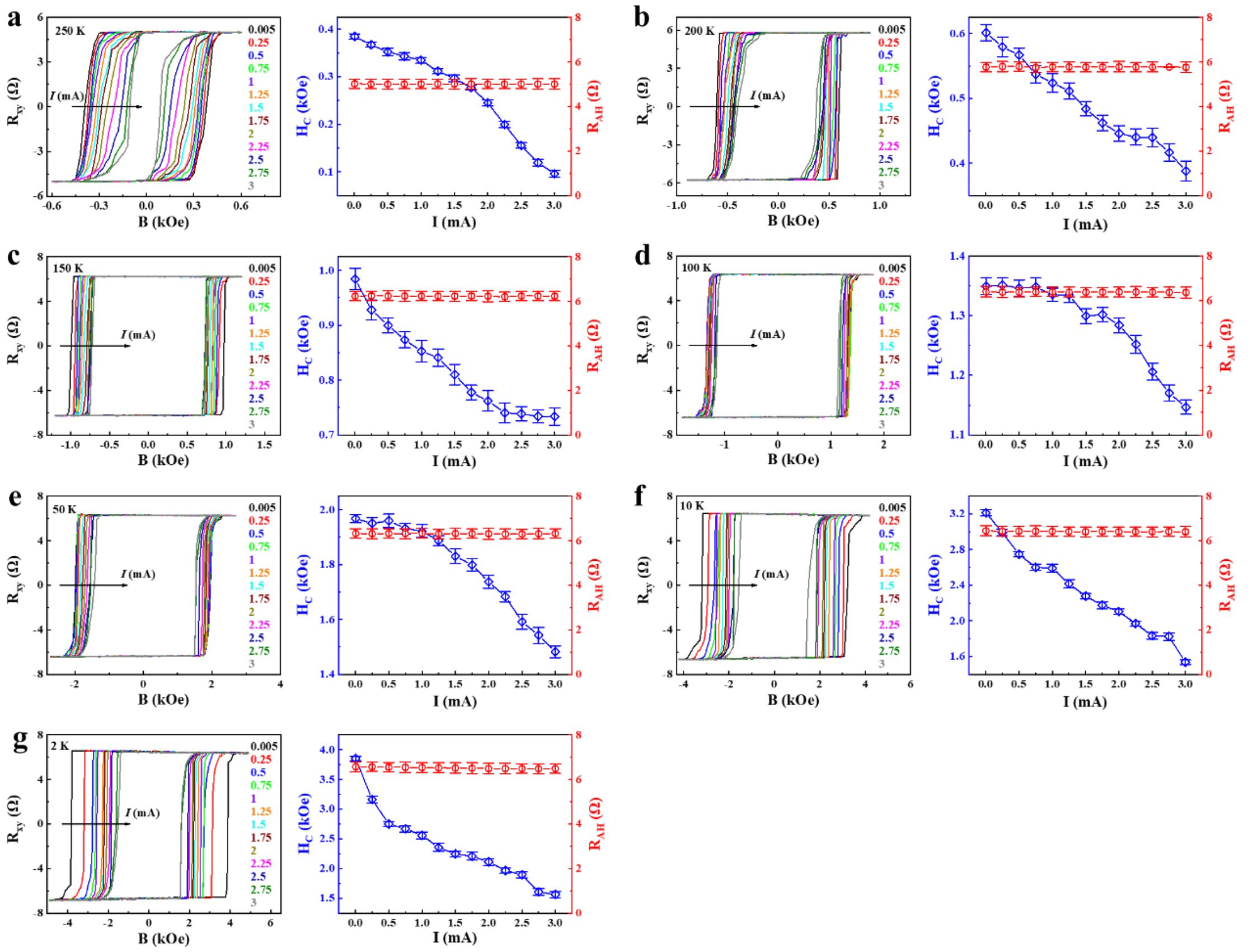

**Figure S6. Current-controlled $H_C$ in 10 nm $Fe_3GaTe_2$ nanosheet under different temperatures. a**, 250 K. **b**, 200 K. **c**, 150 K. **d**, 100 K. **e**, 50 K. **f**, 10 K. **g**, 2 K. Error bars SD, N=3 for $H_C$ and N=25 for $R_{AH}$.

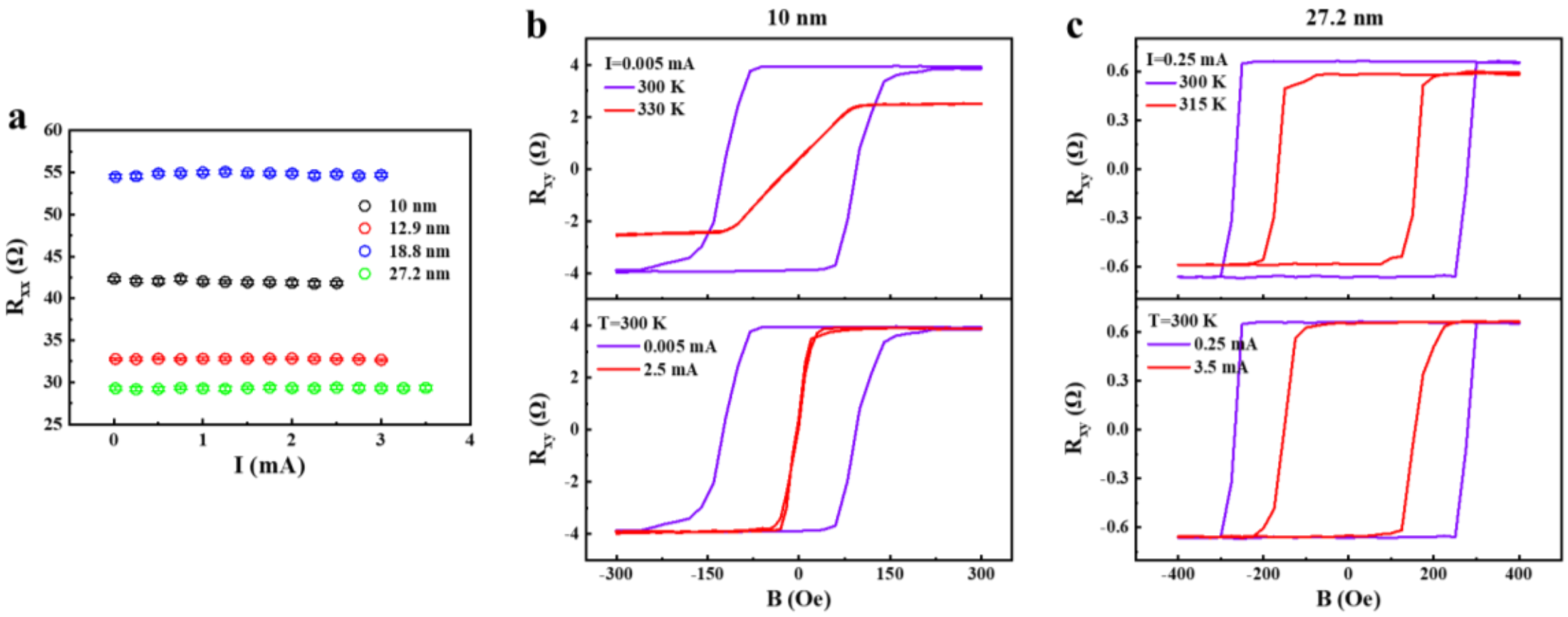

**Figure S7. Elimination of Joule heating effect. a**, Current-dependent $R_{xx}$ in four as-tested 2D $Fe_3GaTe_2$ nanosheets at 300 K. Error bars SD, N=3 for $R_{xx}$. **b**, $R_{xy}$-B curves with similar $H_C$ controlled by temperature (300 and 330 K, I=0.005 mA) and current (0.005 and 2.5 mA, T=300 K) in 10 nm $Fe_3GaTe_2$. **c**, $R_{xy}$-B curves with similar $H_C$ controlled by temperature (300 and 315 K, I=0.25 mA) and current (0.25 and 3.5 mA, T=300 K) in 27.2 nm $Fe_3GaTe_2$.

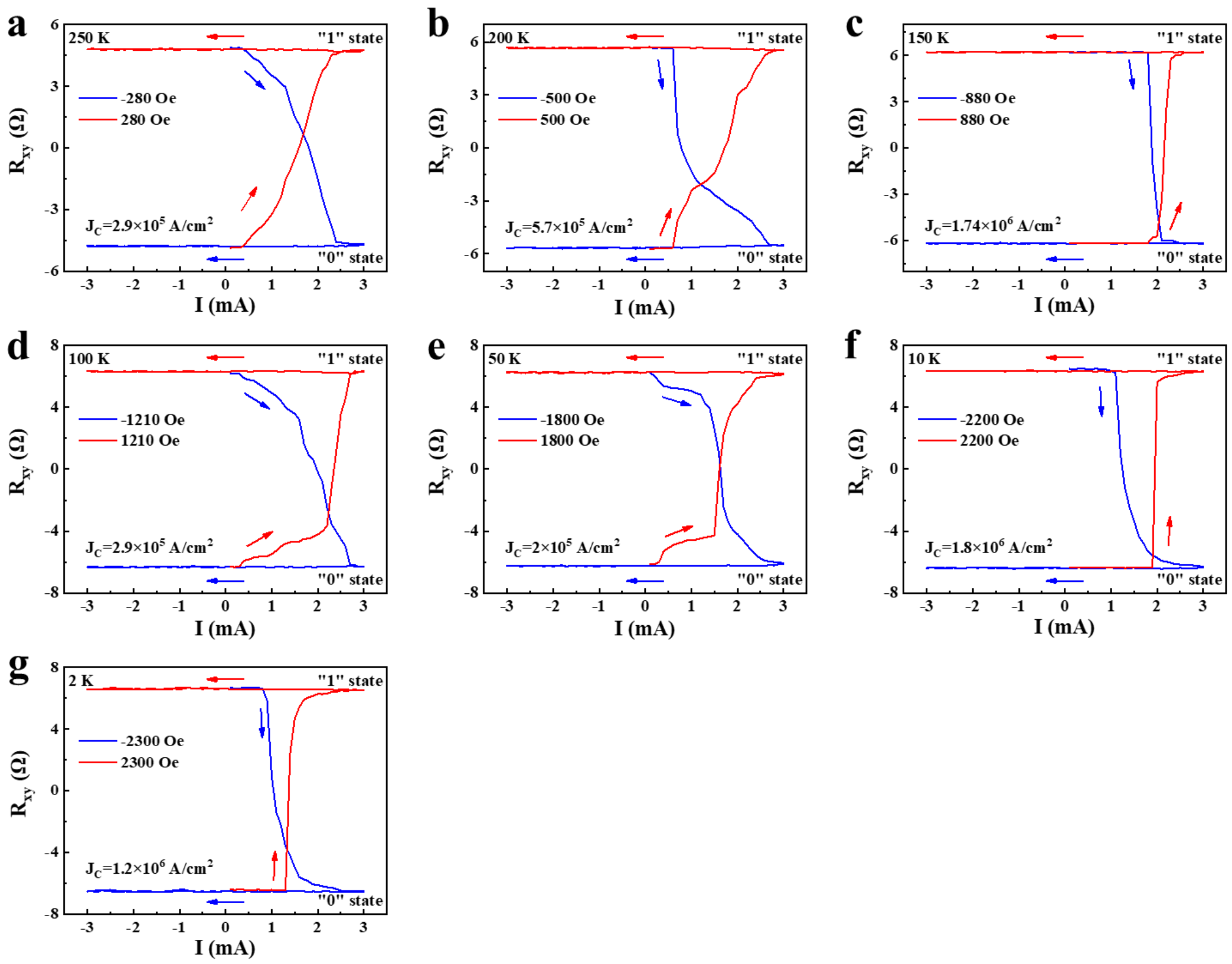

**Figure S8. Current-driven nonvolatile magnetization switching in the 10 nm $Fe_3GaTe_2$ nanosheet under different temperatures. a**, 250 K. **b**, 200 K. **c**, 150 K. **d**, 100 K. **e**, 50 K. **f**, 10 K. **g**, 2 K. The $J_C$ is calculated from the red line in each image.

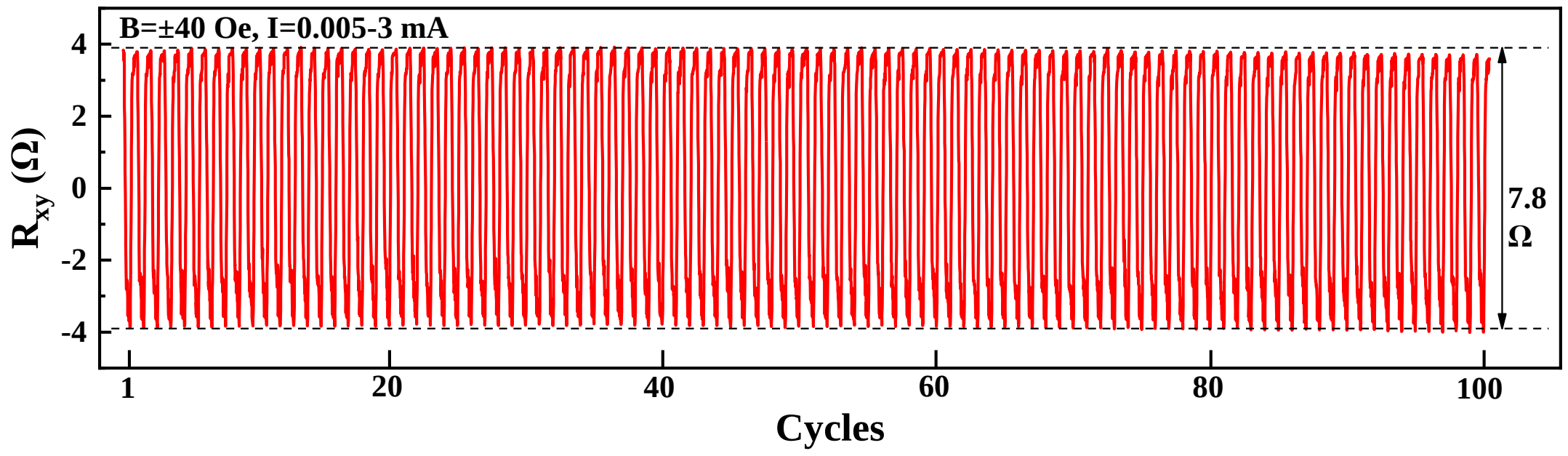

**Figure S9. Cyclic stability of ≈100% magnetization switching of 10 nm $Fe_3GaTe_2$ after 100 times cycles at room temperature.**

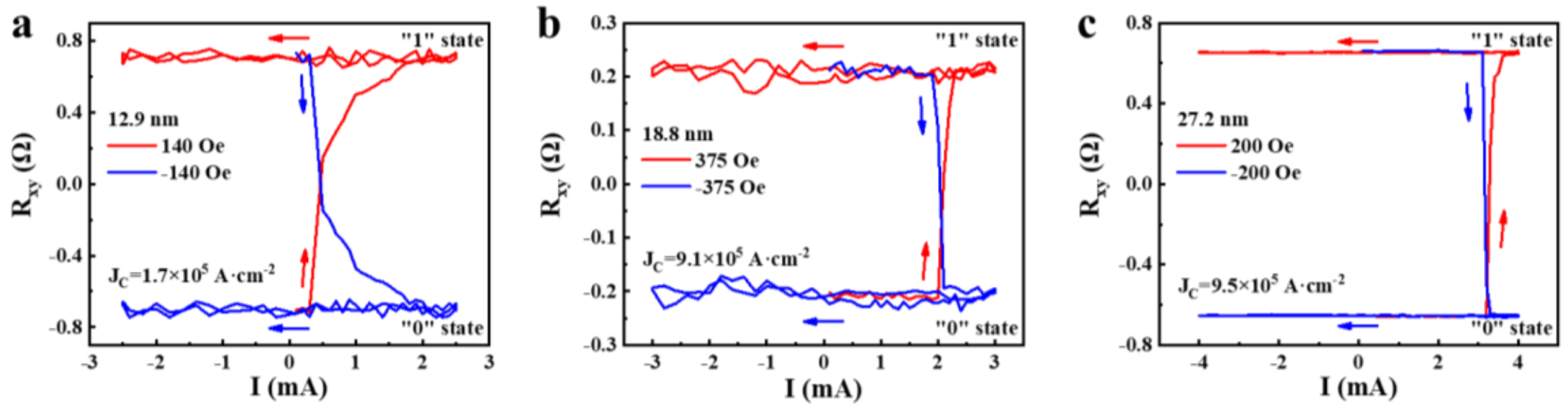

**Figure S10. Room-temperature current-driven nonvolatile magnetization switching in 2D $Fe_3GaTe_2$ nanosheets with different thickness. a,** 12.9 nm. **b**, 18.8 nm. **c**, 27.2 nm. The $J_C$ is calculated from the red line in each image.

**Table S1. Comparison of working temperature, required electric field, and tunable ratio of $H_C$ with various ferromagnetic systems.**

| Structures | Materials | T (K) | E ($mV \cdot nm^{-1}$) | $\Delta H_C/H_C \times 100\%$ | Ref. |
|---|---|---|---|---|---|
| Non-vdW | Co/Pt | 286.8 | 384 | 61.4 | [5] |
| | | 290 | 240 | 25.6 | [6] |
| | (Ga,Mn)As | 35 | 500 | 39 | [7] |
| | Fe/$BaTiO_3$ | 300 | 1 | 30 | [8] |
| | | 150 | 4 | 24 | [9] |
| | $Mn_4N/SrTiO_3$ | 300 | 626 | 2.7 | [10] |
| VdW | $Fe_3GeTe_2$ | 2 | 0.3 | 100 | [11] |
| | $CrI_3$ | 15 | 175.44 | 16.7 | [12] |
| | | 4 | 1600 | 37.5 | [13] |
| | | 6 | 2400 | 30 | [14] |
| | **$Fe_3GaTe_2$** | **300** | **0.029** | **98.06** | **This work** |

**Table S2. Comparison of working temperature and $J_C$ of magnetization switching with various ferromagnetic systems.**

| Structures | Materials | T (K) | $J_C$ (A·cm$^{-2}$) | Ref. |
|---|---|---|---|---|
| Non-vdW ferromagnets | Pt/CoFeB/MgO | 300 | $1.6\times10^7$ | [15] |
| | W/CoFeB/MgO | 300 | $1.6\times10^6$ | [16] |
| | Ta/CoFeB/MgO | 300 | $3\times10^6$ | [17] |
| | $AlO_x$/Co/Pt | 300 | $7\times10^7$ | [18] |
| | IrMn/Co/Ru/CoPt | 300 | $2.8\times10^7$ | [19] |
| | $Bi_{0.9}Sb_{0.1}$/MnGa | 300 | $1.5\times10^6$ | [20] |
| | Pt/$Tm_3Fe_5O_{12}$ | 300 | $1.7\times10^7$ | [21] |
| | $WTe_2$/Py | 300 | $2.96\times10^5$ | [22] |
| | 1T'-$MoTe_2$/Py | 300 | $6.71\times10^5$ | [23] |
| | $Bi_2Se_3$/Py | 300 | $6\times10^5$ | [24] |
| | $Bi_2Se_3$/NiO/Py | 300 | $1.27\times10^7$ | [25] |
| | Pt/NiO/$Y_3Fe_5O_{12}$ | 150 | $3.6\times10^7$ | [26] |
| | $SrRuO_3$/NiO/$SrIrO_3$ | 60 | $8.1\times10^5$ | [27] |
| VdW ferromagnets | Pt/$Fe_3GeTe_2$ | 180 | $2.5\times10^7$ | [28] |
| | $WTe_2$/$Fe_3GeTe_2$ | 150 | $3.9\times10^6$ | [29] |
| | | 150 | $9.8\times10^6$ | [30] |
| | Ta/$Cr_2Ge_2Te_6$ | 10 | $5\times10^5$ | [31] |
| | $ZrTe_2$/$CrTe_2$ | 50 | $1.8\times10^7$ | [32] |
| | $Fe_3GeTe_2$ | 2 | $5\times10^5$ | [11] |
| | **$Fe_3GaTe_2$** | **300** | **$1.7\times10^5$ $1.9\times10^5$** | **This work** |

## Supplementary References